# Annealing driven positive and negative exchange bias in Fe-Cu-Pt heterostructures at room temperature

M. A. Basha[1, 2], Harsh Bhatt[1], Yogesh Kumar[1], C. L. Prajapat[2,3], M. Gupta[4], V. Karki[5], S. Basu[1,2] and Surendra Singh[1, 2,*]

[1]Solid Sate Physics Division, Bhabha Atomic Research Centre, Mumbai 400085, India
[2]Homi Bhabha National Institute, Anushaktinagar, Mumbai 400094, India
[3]Technical Physics Division, Bhabha Atomic Research Centre, Mumbai 400085, India
[4]UGC DAE Consortium for Scientific Research, University Campus, Khandwa Road, Indore 452017, India
[5]Fuel Chemistry Division, Bhabha Atomic Research Centre, Mumbai 400085, India

*surendra@barc.gov.in

**Abstract:**

We report annealing induced exchange bias in Fe-Cu-Pt based heterostructures with Cu as an intermediate layer (Fe/Cu/Pt heterostructure) and capping layer (Fe/Pt/Cu heterostructure). Exchange bias observed at room temperature (300 K) is found to be dependent on the annealing temperature. We obtained positive exchange bias of ~ +120 Oe on annealing both the heterostructures at 400 $^o$C, while on annealing these heterostructures at 500 and 600 $^o$C a negative exchange bias of ~ -100 Oe was found. X-ray reflectivity and polarized neutron reflectivity measurements provided evolution of depth dependent structure and magnetic properties of the heterostructures on annealing at different temperatures and revealed coexistence of soft and hard (alloy) magnetic phases across the thickness of the films. Rapid and long range interdiffusion at interfaces on annealing the systems at a temperature > 400 $^o$C resulted into formation of a ternary alloy phase. These results can be understood within the context of a very unusual interface exchange interaction at the interface of hard/soft magnetic phases, which are dependent on the annealing temperature.

**INTRODUCTION**

Ferromagnetic (FM)/ heavy metal systems have been investigated for various spin based phenomena such as spin orbit torques [1], spin hall effect [2, 3], spin pumping [4,5] and the Dzyaloshinskii-Moriya interaction [6,7]. Fe (FM)/Pt (heavy metal) system, in particular, has attracted considerable attention in recent years because of its magnetic properties that are deployable in ultrahigh-density information storage [8-11]. The ordered $L_{10}$ or face centered tetragonal (FCT) FePt phase exhibits very high uniaxial magnetocrystalline anisotropy along the *c* direction of the crystal structure and shows high perpendicular magnetic anisotropy (PMA) which is useful in high density recording media [8-11]. Nanoparticles of FePt ordered alloy have also shown promising electro-catalysis towards the oxygen reduction reaction in a proton-exchange membrane fuel cell [12-15]. However, to obtain ordered FePt alloy (hard magnetic) film from as-deposited disordered (FCC phase) FePt films, annealing of the disordered films after deposition is required. In general, an annealing temperature of $T_a > 600°C$ is required to transform the disordered structure to the ordered structure.

Recently, efforts have been made to reduce the ordering temperature by introducing (a) a buffer layer [16-20], (b) a cap layer [21, 22], and (c) addition of third element to form a ternary alloy compound [23-26]. One of the effective ways to reduce the ordering (FCC to FCT) temperature of FePt film is by adding Cu to FePt and realizing the formation of $(FePt)_{1-x}Cu_x$ ternary alloys (x = 0-30%) [27-32]. These ternary alloys show drastically different magnetic properties [27-32]. Earlier, we studied the formation of ternary alloy FePtCu on annealing FePt/Cu multilayer, where Cu thickness was larger than the FePt alloy and it resulted in formation of $(FePt)_{0.42}Cu_{0.58}$ [26]. The ternary alloy of FCT phase was obtained on annealing the multilayer at 400 °C, due to high diffusivity of the elements in this system, which exhibited a superparamagnetic behavior.

The exchange bias effect at room temperature [33,34] in a coupled FM layer system without any field cooling procedure also generates a lot of interest. Usually, exchange bias is a consequence of the interfacial exchange interaction between a FM and an antiferromagnetic (AFM) material [33,34] and is manifested by a shift of the FM hysteresis loop along the magnetic field axis after field cooling through the Neel temperature of the AFM. An induced exchange bias behavior has also been observed in systems with two coupled FM materials such as a Pt/Co multilayer and a NiFe thin film [35,36], which is attributed to the interplay between out-of-plane and in-plane anisotropies of the Pt/Co multilayer and NiFe thin film, respectively. The composite films consisting of $L_{10}$ FePt as hard magnetic and soft magnetic (e.g. Fe, Co etc.) phases also exhibit interface dependent complex magnetic properties, e.g. exchange spring phenomena [37, 38], exchange bias effect [39-42], magnetization reversal process [43, 44] etc. Both disordered and ordered FePt phases are expected to form on thermal annealing of Fe/Pt multilayer, which may give rise to these complex magnetic properties. However, to understand the mechanism and kinetics of the transformation of Fe-Cu-Pt heterostructures due to annealing a systematic interdiffusion and depth dependent structure-magnetic correlation in the nm length scale is highly desirable.

Here we report on annealing induced room temperature exchange bias effect in Fe-Cu-Pt heterostructure grown on Si substrates, the system which in the as-grown state exhibits soft ferromagnetism with easy axis in the plane of the film. The evolution of structure-magnetic properties, of Fe-Cu-Pt heterostructures with Cu as intermediate layer (Fe/Cu/Pt heterostructure) and capping layer (Fe/Pt/Cu heterostructure), as a function of annealing temperature has been studied using a combination of techniques e.g. Grazing incidence X-ray diffraction (GIXRD), Secondary ion mass spectrometry (SIMS), superconducting quantum interference device (SQUID), x-ray reflectivity (XRR) and polarized neutron reflectivity (PNR). Both the heterostructures exhibited positive exchange bias of ~ +120 Oe on annealing at 400 $^{o}$C and a negative exchange bias ~ -100 Oe, on further annealing of the systems at 500 and 600 $^{o}$C, respectively.

**EXPERIMNETAL DETAILS**

Two heterostructures of Fe, Cu and Pt with a nominal structure of Si(substrate)/[Fe(200Å)/Cu(50Å)/Pt(200Å)] and Si(substrate)/[Fe(200Å)/Pt(200Å)/Cu(50Å)], henceforth known as S1 and S2, respectively, were grown on silicon (100) substrates using dc magnetron sputtering. Before deposition, a base pressure of $1\times10^{-5}$ Pa was achieved. The substrate was kept at room temperature during growth of these heterostructure and it was rotated along its own axis at 60 rpm for achieving greater uniformity. The exact individual thickness of each layer was estimated using XRR measurements. Evolution of structure and magnetic properties of the interfaces and alloy formation at interfaces of heterostructures were studied as a function of annealing the heterostructures at 300, 400, 500 and 600 °C under a vacuum same vacuum (~$10^{-5}$ Pa) for a time period of 30 minutes at each annealing stage.

The evolution of crystalline structure on annealing of heterostructures was investigated using grazing incidence x-ray diffraction (GIXRD) with Cu $K_\alpha$ radiations. Elemental depth distribution analysis of Fe, Cu and Pt present in the as-deposited and post annealed at 600 °C were carried out using Cameca IMS-7f secondary ion mass spectrometry (SIMS) instrument equipped with both oxygen duo plasmatron and cesium thermal ion source. $Cs^+$ primary ion beam with a beam current of 10±1 nA and impact energy of 5 keV was raster scanned over an area of 250 µm × 250 µm on the heterostructure surface. Small pieces of the heterostructures were used for magnetization measurements using a superconducting quantum interference device (SQUID). All the magnetization measurements reported in this paper were measured by applying the field along the plane of the film. The depth dependent structural and magnetic properties of the heterostructures were characterized using x-ray reflectivity (XRR) and polarized neutron reflectivity (PNR) [45-49]. PNR experiments were carried out using polarized neutron reflectometer instrument (neutron wavelength ~ 2.9 Å) at DHRUVA, India [45]. An in-plane magnetic field of 1.7 kG was applied on the heterostructures during PNR measurements.

XRR and PNR are two non-destructive complementary techniques to study the depth profiling of chemical and magnetic structure of multilayer heterostructure with a depth resolution of sub nanometer length scale averaged over the lateral dimensions (~ 100 $mm^2$) of the heterostructure [45-49]. The specular reflectivity, $R$, of the heterostructure was measured as a function of wave vector transfer, $Q = 4\pi \sin\theta/\lambda$ (where, $\theta$ is angle of incidence and $\lambda$ is the x-ray or neutron wavelength). The reflectivity is qualitatively related to the Fourier transform of the scattering length density (SLD) depth profile $\rho(z)$, averaged over the whole heterostructure area. In case of PNR, $\rho(z)$ consists of nuclear and magnetic SLDs such that $\rho^\pm(z) = \rho_n(z) \pm \rho_m(z)$. The +(-) sign denotes neutron beam polarization along (opposite to) the applied field and corresponding reflectivities are denoted as $R^\pm$. The layer structure were obtained from the XRR data by fitting model electron SLD (ESLD) profiles, $\rho(z)$ that fit the reflectivity data. The reflectivities were calculated using the dynamical formalism of Parratt [48] and parameters of the model (thickness and ESLD) were fitted using a genetic algorithm based program [49], which adjust the parameter to minimize the value of reduced $\chi^2$ –a weighted measure for goodness of fit [50].

**RESULTS AND DISCUSSION**

*GIXRD measurements:*

Fig. 1 (a) and (b) show the GIXRD data at different annealing temperature recorded for S1 and S2 heterostructures. We observed polycrystalline *fcc* and *bcc* structure [51] for Pt and Fe layers, respectively, for both the as-deposited heterostructures. However Pt layer in S1 and S2 show preferential growth along (111) and (200) direction, respectively, this can be seen in Fig. 1(c) where we have plotted the intensity ratio of (111) and (200) peaks ($I_{Pt(111)}/I_{Pt(200)}$) as a function of annealing temperature (0 °C corresponds to as-deposited heterostructures). We didn't observe any change in crystalline structure on annealing the heterostructures at 300 °C, except that there was a small change in the intensity ratio ($I_{Pt(111)}/I_{Pt(200)}$), which decreases (increases) on annealing S1(S2) heterostructure at 300 °C. GIXRD data from S1 annealed at 400 °C didn't show much difference and only a reduction in intensity ratio (Fig.1(c)) was observed. In contrast, on annealing S2 at 400 °C we observed additional Bragg peaks correspond to ordered (*fct*) FePt alloy phase [51] (Inset shows the double peaks correspond to Pt(111) and FePt(111) near 2θ ~40 degree, for S2 annealed at 400 °C). Formation of thin *fct* alloy layer on annealing heterostructure S2 at 400 °C as compared to S1 clearly indicate low interdiffusion at interfaces in these nm length scales at low temperatures ≤ 400 °C. For S2, Pt is directly grown on Fe as compared to S1 where thin Cu layer separates Fe and Pt layers therefore FePt alloy formed in S2 is due to low interdiffusion of Fe and Pt at 400 °C. with further annealing of both the heterostructures at 500 and 600 °C, we observed formation of a polycrystalline FePtCu ternary alloy, emerging of (001) and (002) peaks also confirms the same. Bragg peaks in GIXRD data for heterostructures annealed at 500 and 600 °C are indexed in Fig. 1(a) suggesting ordered (*fct*) ternary alloy phase of FePtCu with a preferential growth along (101) direction [51].

Evolution of structural properties as a function of annealing temperature is further studied by estimating the crystallite size of different elements in heterostructures. The Scherrer formula: $t = \frac{0.9\lambda}{B\cos\theta}$, which relates the crystallite size *t* to the angular broadening *B* (in radians) at the Bragg reflection $\theta$ and x-ray wavelength $\lambda$, was used to estimate the crystallite size. We have used high intensity Bragg peaks (reflections (111) and (200) of Pt in heterostructure S1 and S2, respectively upto annealing temperature $\leq$ 400 $^o$C, and reflection (100) for ternary alloy from both heterostructures at annealing temperature $\geq$ 500 $^o$C) to calculate the crystallite size. Fig. 1 (d) shows the variation of crystallite size as a function of annealing temperature. As-deposited heterostructures S1 and S2 show a grain size of ~ 10.7 nm and ~ 9 nm, respectively for Pt. On annealing heterostructures at 300 and 400 °C a small decrease in the grain size was observed. Ternary FePtCu alloy phase with a crystallite size of ~ 10 nm was observed on annealing the heterostructures at 500 °C. A small reduction in grain size of ternary alloy is obtained on further annealing of heterostructures at 600 °C. Thus post annealing above 400 $^o$C, fast, long-range interdiffusion is observed which accompanied evolution of polycrystalline *fct* FePtCu alloy phase for both the heterostructures.

*Macroscopic magnetization measurements (SQUID)*

Fig. 2 (a) and (b) show the room temperature (RT ~ 300 K) magnetic hysteresis curves at different annealing temperatures for S1 and S2 heterostructures (with similar lateral dimensions). Magnetic hysteresis loops were measured after saturating the in plane magnetization of the heterostructure in an applied field of ~ +2500 Oe. S2 shows smaller saturation magnetization ($M_s$) as compared to that of S1 for as-deposited heterostructures. We didn't observe any significant change in macroscopic magnetization properties of both the heterostructures on annealing at 300 °C. Observation of very small coercive field ($H_c \approx$ 30 Oe) for as-deposited S1 and S2 as well as heterostructures annealed at 300 °C suggests soft ferromagnetic nature for both the heterostructures. Variation of the ratio of remnant magnetization ($M_r$) with $M_s$ ($M_r/M_s$) and $H_c$ as a function of annealing temperature have been shown in Fig. 2 (c) and (d), respectively. Annealing of heterostructures at 400 °C indicates sharp increase (decrease) in coercivity (saturation magnetization) and we obtained $H_c$ ~ 310 Oe for S1 and 175 Oe for S2, which is ~ 10 times of the $H_c$ for as-deposited heterostructures. Interestingly, we also observed a shift in hysteresis curve towards positive field (equivalent to positive exchange bias ($E_B$)) axis with an $E_B \approx$ +120 Oe for both the heterostructures. On annealing the heterostructures at 500 °C, we obtained a small increase in $H_c$, which decreases further on annealing the heterostructures at 600

°C (Fig. 2(d)). Remarkably we observed another shift in hysteresis loop ($M$ ($H$)) on annealing the heterostructures at 500 and 600 °C, which is towards the negative field axis (equivalent to negative exchange bias ($E_B$)). Thus we obtained an $E_B$ of ~ -100 Oe on annealing the heterostructures S1 and S2, respectively, at these higher temperatures, which is in contrast to $E_B$ (~ + 120 Oe) obtained for heterostructures annealed at 400 °C. Fig. 2(e) shows the variation of the $E_B$ as a function of annealing temperature. Fig. 2 (c) clearly suggested a reduction in $M_r/M_s$ of both the heterostructures annealed at temperatures ≥ 400 °C. The $M_r/M_s$ ratio decreases from 0.95 (soft FM) to 0.20 (hard FM) on annealing the heterostructures from 300 to 600 °C, suggesting the modification in magnetization of whole system. Occurrence of exchange bias phenomenon in these heterostructures on annealing above 300 °C indicate the presence of plausible hard and soft magnetic phases in the system, where Fe and alloy layer (e.g. FePt, FePtCu, FeCu etc.) act as a soft and hard ferromagnetic layer, respectively. Previous reports also suggested observation of $E_B$ in FePt/Fe systems [39-42].

Temperature dependent macroscopic magnetization properties of the heterostructures S1 and S2 at different annealing temperature are shown in Fig. 3 (a) and (b) respectively. We have measured field cooled (FC) and zero field cooled (ZFC) magnetization data ($M$ ($T$)) as a function temperature after cooling the heterostructure from 300 K to 5 K in an applied in-plane magnetic field of 500 Oe (FC) and 0 Oe (ZFC). FC and ZFC data were collected while warming the heterostructure in a field of 500 Oe. The FC and ZFC data as a function of temperature for as-deposited heterostructures and heterostructures annealed at 300 °C remain same throughout the temperature range of measurements. The temperature dependent variation of difference between FC and ZFC magnetization data from heterostructures S1 and S2 annealed at different temperatures 400 °C - 600 °C, are shown in Fig. 3(c) and (d), respectively. An irreversible magnetic behavior (difference between FC and ZFC) is evident, for both the heterostructures annealed at 400 °C, in the whole range of temperature; this indicates the presence of a non negligible fraction of nanoparticles which are blocked above/near RT so that the systems show hard magnetic properties even at RT. On annealing the heterostructures at higher temperatures (500 and 600 °C) the irreversibility of FC and ZFC data slightly shift below RT (Fig. 3 (c) and (d)). This is also reflected from $H_c$ values obtained for these systems on annealing at higher temperatures (Fig. 2(d)). However the decrease was predominant in case of S2 as compared to that of S1. We believe the macroscopic magnetization modulations of these systems are due to formation of alloy phases at interfaces and hence depth dependent structure-magnetic property investigation will help to understand these phenomena.

*Depth dependent structure: XRR measurements*

Fig. 4(a) and (b) show the XRR data from S1 and S2, respectively, for as-deposited and post annealed states. XRR data for different conditions are shifted vertically for better visualization. Continuous lines in Fig. 4(a) and (b) represent the corresponding fit to XRR data for different annealing temperatures. Fig. 4 (c) and (d) show the ESLD depth profile of heterostructures S1 and S2, respectively, for as-deposited and post annealing at different temperatures. The parameters extracted from XRR measurements for as-deposited heterostructures S1 and S2 are given in Table 1. However small variation (within the error on parameters) in ESLD along the thickness of the Fe and Pt layers were considered to get best fit to XRR data (Fig. 4(c) and (d)).

On annealing of heterostructure S1 at 300 °C, we didn't observe any significant change in the depth dependent ESLD. However on annealing S2 at 300 °C a small variation of ESLD especially at interfaces was observed, suggesting interface dependent interdiffusion of elements (for S2, Fe and Pt diffuses fast as they are in direct contact with each other). On annealing the heterostructures at 400 °C we observed formation of alloy phases at interfaces. GIXRD data clearly suggested formation of crystalline FePt alloy on annealing the heterostructure S2 at 400 °C and same can be observed as a change in ESLD (~ $1.04 \times 10^{-5}$ Å$^{-2}$ for alloy layer) at the Fe/Pt (Pt on Fe interface) interface from XRR measurements (Fig. 4 (d) highlighted area). There is also a change in ESLD at Pt/Cu interface, suggesting interdiffusion of Cu and Pt in heterostructure. We also observed a variation in ESLD at the Cu/Pt and Fe/Cu interfaces in the heterostructure S1 on annealing at 400 °C. However GIXRD data did not show any crystalline phase for the alloy phases in S1 on annealing at 400 °C. with further annealing of both the heterostructures at 500 and 600 °C, we observed a drastic change in depth dependent layer structures (XRR data) which precisely corroborate with crystalline structure obtained from GIXRD measurements. We obtained a single alloy layer formation with an ESLD of ~$(9.6 \pm 0.5) \times 10^{-5}$ Å$^{-2}$ (corresponds to a ternary alloy FePtCu) for both the heterostructure on annealing at 500 and 600 °C, suggesting high interdiffusion of elements at interfaces (Fe/FePt, Pt/FePt, Cu/CuPt etc., formed on annealing at 400 $^{o}$C) on annealing at 500 $^{o}$C. We have observed small reduction in total thickness of heterostructure S1 and S2 on annealing at 500 and 600 $^{o}$C.

*SIMS Measurements*

In order to see the elemental distribution along the depth of the heterostructures we have also performed SIMS measurements for as-deposited heterostructures and post annealed heterostructures at 600 $^{o}$C. SIMS data from as-deposited heterostructures S1 and S2 are shown in Fig. 5(a) and (b), respectively. Fig. 5(c) and (d) show the SIMS data from heterostructures S1 and S2, respectively, on annealing at 600 $^{o}$C. The time-axis represents the depth and the intensity-axis represents the concentration. The Fe/Si interface (film-substrate interface) is indicated by vertical dashed lines in Fig. 5(a-d). We observed nearly constant intensity of Pt, Cu and Fe for different regions along the depth (similar to their growth sequence) up to the Fe/Si interface, signifying the homogeneity of different layers in as-deposited (as-dep) heterostructures (Fig. 5(a) and (b)) and consistent with XRR measurements. It is evident from Fig. 5(c) and (d) that Pt, Cu and Fe show nearly constant intensity for whole range of the film up to the Fe/Si interface, indicating formation of a homogeneous layer (alloy layer) on annealing of both the heterostructures S1 and S2 at 600 $^{o}$C. SIMS data from heterostructure S1 annealed at 600 $^{o}$C also showed that interface corresponding to Si/Fe interface (vertical line in Fig. 5(c)) is reached faster then as-deposited heterostructures (Fig. 5(a-b)) and heterostructure S2 annealed at 600 $^{o}$C (Fig. 5(d)), suggesting reduction in total thickness of S1 on annealing at 600 $^{o}$C. Thus the SIMS data for as-deposited and post annealed heterostructures at 600 $^{o}$C corroborated the depth dependent structure obtained from XRR measurements.

*Depth dependent structure and magnetic properties: PNR measurements*

Fig. 6 shows the PNR measurements from as-deposited and post annealed heterostructures at different temperatures. Closed circles and triangles in Fig. 6(a-j) depict the spin up ($R^+$) and spin down ($R^-$) PNR data and continuous lines are corresponding fit to PNR data. In general the difference between spin dependent reflectivities ($R^+ - R^-$) provides the magnetization depth profile in the heterostructures. Modifications in spin dependent PNR data at different annealing temperature are evident from Fig. 6, which clearly suggest different evolution process of magnetization on annealing of the heterostructures. Fig. 7 (a) and (b) show the NSLD depth profile along with the corresponding MSLD (blue shaded regions) depth profile across the interfaces of heterostructures S1 and S2, respectively, at different annealing temperatures. These NSLD and magnetization profiles were obtained from the best fit of corresponding PNR data at different annealing temperatures. The structural parameters for as-deposited heterostructures obtained from PNR data are listed in the Table 1, along with the parameters obtained from XRR data and both are found to be consistent with each other.

We obtained an MSLD of ~ $(4.40\pm0.35)\times10^{-6}$ Å$^{-2}$ (average magnetic moment of ~$1.80\pm0.12$ $\mu_B$/atom) for as-deposited heterostructures S1 and S2. On annealing the heterostructures S1 and S2, at 300 °C, a small decrease in MSLD ~ $(3.85\pm0.45)\times10^{-6}$ Å$^{-2}$ (~$1.62\pm0.12$ $\mu_B$/atom) was observed, which may be due to increase in inhomogeneties of Fe layer due to interdiffusion at this temperature. Further decrease in magnetization is observed for both the heterostructures on annealing at 400 °C. Larger reduction in magnetization for heterostructure S1 on annealing at 400 °C was observed suggesting formation of possible alloy phase/mixing at Fe/Cu and Cu/Pt interfaces (highlighted by vertical lines in Fig. 7(a)) with small magnetic contribution. The different magnetic response of the alloy phase (may be hard magnetic properties) and rest of Fe (soft phase) layer on annealing of S1 at 400 °C may also be contributing to the possible positive exchange bias in this heterostructure. While annealing of heterostructure S2 at 400 °C we also observed formation of alloy at Fe/Pt interface but it is FePt phase as seen in GIXRD pattern, thus the hard/soft (FePt/Fe) interface in this system contribute to positive exchange bias. In addition S2 shows a small reduction in magnetization on annealing it from 300 °C to 400 °C as compared to that of S1, which further confirm the FM phase of FePt alloy in S2. Further reduction/modification in magnetization of the heterostructures post annealing at 500 and 600 °C was observed. Post annealing at 500 and 600 °C we obtained a ternary alloy formation. Small variation in ESLD (from XRR) and NSLD (from PNR, Fig. 7(a) and (b)) of S1 and S2 post annealing at 500 and 600 °C, exhibit the coexistence of different phases (hard and soft magnetic) along the thickness of the film, which may results into low magnetization and a negative exchange bias. The saturation magnetization at room temperature obtained by SQUID and thickness weighted magnetization obtained from PNR as a function of annealing temperature for heterostructures S1 and S2 are shown in Fig. 8 (a) and (b) respectively, suggesting macroscopic magnetization (SQUID) measurements are consistent with the PNR measurements.

The composition of the alloy layer formed on annealing of a heterostructure as a result of interdiffusion at interfaces and complete mixing can be theoretically calculated using their density and thickness [46,52]. For binary alloy system with two element (say A and B), the composition ratio (x : y) for alloy can be calculated using $x/y = n(A)d(A)/n(B)d(B)$, where $n(A)$ and $n(B)$ are density of A and B, respectively and $d(A)$ and $d(B)$ are the thickness of these layers. Using the density of Fe, Pt and Cu elements, the particle density of these elements will be $n(Fe) = 8.47 \times 10^{22}$ cm$^{-3}$, $n(Pt) = 6.61 \times 10^{22}$ cm$^{-3}$ and $n(Cu) = 8.49 \times 10^{22}$ cm$^{-3}$, respectively. Within first approximation, using the layer thickness and number (particle) density of each element, we have calculated the expected composition of ternary alloy phase, which are $(Fe_{0.56}Pt_{0.44})_{0.9}Cu_{0.1}$ and $(Fe_{0.61}Pt_{0.39})_{0.9}Cu_{0.1}$ for heterostructures S1 and S2, respectively. The theoretical ESLD and NSLD for these alloy phases in heterostructure S1 (S2) are $10.1 \times 10^{-5}$ Å$^{-2}$ ($10.2 \times 10^{-5}$ Å$^{-2}$) and $6.58 \times 10^{-6}$ Å$^{-2}$ ($7.0 \times 10^{-6}$ Å$^{-2}$), respectively. The ESLD and NSLD for the alloy phases in heterostructure S1 (S2) (on annealing at 500 and 600 $^\circ$C) obtained from XRR and PNR were $9.6 \times 10^{-5}$ Å$^{-2}$ ($9.5 \times 10^{-5}$ Å$^{-2}$) and $6.2 \times 10^{-6}$ Å$^{-2}$ ($5.7 \times 10^{-6}$ Å$^{-2}$), which are in close agreement to the theoretically calculated values for these alloy phases and hence confirmed the ternary alloy phase formation on annealing the heterostructures at 500 and 600 $^\circ$C.

The $E_B$ in a ferromagnetic/antiferromagnetic system is attributed to a competition between the interfacial exchange interaction and the Zeeman energy [33, 34]. $E_B$ has also been reported in systems with ferrimagnetic/ferromagnetic [53], ferrimagnetic/antiferromagnetic [54], ferrimagnetic/ferrimagnetic [55, 56], and hard/soft ferromagnetic/ferromagnetic interfaces [39-42]. In general, $E_B$ is a measure of shift of hysteresis loop opposite to the cooling field, i.e., negative $E_B$. In some systems, positive $E_B$ were found, where the shift is in the same direction as the cooling fields. The exchange bias field, $E_B$ is given by [53]: $E_B = \Delta\varepsilon / 2M_f t_f$, where $M_f$ is the saturation magnetization of the ferromagnetic layer and $t_f$ is its thickness. $\Delta\varepsilon$ is difference between the interface energies for a net magnetization of the ferromagnetic layer in parallel and antiparallel direction to the applied magnetic field. Depending on coupling at the interface, $\Delta\varepsilon$ will be positive (antiferromagnetic) or negative (ferromagnetic), which will determine the type of exchange bias. The schematics of two types of coupling and related exchange bias phenomena are depicted in Fig. 8 (c) and (d).

Evolution of structure and magnetic properties of Fe-Cu-Pt heterostructures as a function of annealing temperature clearly suggested the formation of alloy phase above 400 $^0$C, which also showed modification in magnetization of the system. XRR and PNR results also suggested minor inhomogeneties in scattering length density depth profiles for alloy phase on annealing of both the heterostructure above 400 $^0$C, which may suggest coexistence of different phases (e.g. Fe, FePt, FePtCu, FePt (*fcc* and *fct*). However ordered *fct* phase of alloy was in majority on annealing the heterostructures above 400 $^0$C, which was also confirmed by GIXRD measurements. Coexistence of these phases provides a matrix of soft/superparamagnetic (Fe / *fcc* alloy phases of FePt and FePtCu) and hard (*fct* alloy phases of FePt and FePtCu) ferromagnetic phases. Post annealing of both the heterostructures at 400 $^o$C exhibited a shift of hysteresis loop to positive field direction (positive exchange bias) at RT, where hard phase (FePt, FePtCu etc.) are in minority. Coexistence of soft-hard magnetic phases on annealing of the heterostructures at temperatures ≥ 400 $^o$C may also be contributing to magnetic disorder in the system which results in to the splitting of FC-ZFC magnetization data for these annealing temperatures. The coexistence of hard and soft magnetic phases along the thickness may create magnetic domains at the interface of these phases that differ from the rest of the film (majority of phase) and hence may be contributing to exchange bias in this system. Reduction in saturation magnetization on annealing ≥ 400 $^o$C for both the heterostructures also suggests different magnetic phase (antiferromagnetic/superparamagnetic etc.) at interfaces of the hard-soft magnetic phases. Positive and negative exchange bias has already been observed in hard/soft magnetic core/shell nanoparticles [57], bulk manganite $NdMnO_3$ [58]. Aas et al.,[59], in their theoretical investigation, have observed a modification in the exchange interaction at the interfaces of Fe/FePt (soft/hard magnet) system and observed an small antiferromagnetic coupling at the interfaces. We propose that the combination of ferromagnetic and antiferromagnetic coupling at soft/hard magnet interfaces in present system and annealing induced domination of these phases might be contributing to the presence of both positive and negative exchange bias in present system at different annealing temperatures.

These results confirm not only that $L_{10}$ ordering of ternary alloy occurs, but also that inhomogeneties of other phases also formed upon annealing, which modulate the magnetic properties. However earlier reports [60-62] also suggested diffusion of Fe in Si substrate at ~ 800 $^o$C , which leads to formation of iron silicide and a strong decrease of the magnetic moment value to 0.97 µB/atom at low temperatures. Our measurements are well below the above mentioned annealing temperature, so such a contribution is negligible. However to see any such effect we have simulated PNR data by considering a thin (thickness ~ 6 nm) FeSi$_2$ layer at Si/Fe interface (substrate/film interface) with a magnetic moment of 0.97 µB/atom. Fig. 9 (a) show the normalized spin asymmetry (NSA = (R$^+$ - R$^-$)/(R$^+$ + R$^-$), where R$^\pm$ are spin dependent PNR data), data for heterostructure S2 annealed at 600 °C and fits assuming: (a) a single alloy layer (Fig. 9 (b)) which best fitted the data and discussed earlier (Fig. 6 and 7), (b) considering alloy layer and a FeSi$_2$ layer (thickness ~ 6 nm) at Si/Fe interface (Fig. 9 (c)). The fit for NSA data assuming model (a) and (b) are shown as solid line (blue) and a line with star (black), respectively, in Fig. 9 (a). It is clear from Fig. 9 that considering thin FeSi$_2$ layer with a magnetic moment of 0.97 µB/atom did not fit the NSA (PNR) data over whole Q-range.

Further to see the behavior of exchange bias at low temperature we compared the $M(H)$ curve (Fig. 10) at 300 and 5 K, for as-deposited and post annealed heterostructures S1 and S2 at different temperatures. It is evident from Fig. 10 that we observed similar exchange bias phenomena at 300 K and 5 K upon post annealing of both the heterostructure above 300 °C. For as-deposited and annealed heterostructures S1 and S2 at 300 °C, we obtained increase in $H_C$ at 5K, which is well known phenomena for soft Fe film. The other features of hysteresis curve e.g. $M_s$ and $M_r/M_s$, remains same at low temperature on annealing the heterostructures at different temperatures, suggesting good temperature-dependent thermal stability for the observed magnetic properties of the systems on annealing at different temperatures.

**CONCLUSION**

In summary, annealing induced structure and magnetic properties of Fe-Cu-Pt heterostructures with Cu as a capping and intermediate layer has been studied. A very small and short range interdiffusion of elements (Fe, Cu and Pt) across the interfaces was observed in heterostructures upon annealing at temperature ≤ 400 °C. Whereas a rapid and long range interdiffusion with alloy formation upon annealing of the heterostructures at temperature > 400 °C was observed. Both the heterostructures exhibited a positive exchange bias (~ +120 Oe) at room temperature and 5 K, upon annealing at 400 °C. However annealing of the heterostructures at a temperature > 400 °C resulted into negative exchange bias (~ -100 Oe) at room temperature and 5 K. Depth dependent structure and magnetic properties obtained from XRR and PNR suggested strong correlation between the transition in exchange bias for heterostructures on annealing at and above 400 $^o$C with coexistence of hard-soft magnetic phases along the thickness of the systems. The observation of annealing induced exchange bias effect suggest different sign of the interface magnetic exchange interaction, which is highly dependent on the majority of phase (hard/soft), present in the system and hence can prove useful in magnetic devices.

## ACKNOWLEDGEMENT

Authors acknowledge the help of Nidhi Pandey and Prabhat Kumar of UGC-DAE-CSR Indore centre during deposition of the films and Swapan Jana, V. B. Jayakrishnan of Solid State Physics Division, BARC during X-ray scattering measurements of the films. One of the authors, Yogesh kumar, would like to acknowledge Department of Science and Technology (DST), India for financial support *via* the DST INSPIRE Faculty research grant (DST/INSPIRE/04/2015/002938).

**Table 1**: Parameters for Fe-Pt-Cu heterostructures S1 and S2 obtained from XRR and PNR for as-deposited condition. Parameters shown in brackets are those obtained from PNR measurements.

| **Parameters** | **S1: Si/Fe/Cu/Pt** | | | **S2: Si/Fe/Pt/Cu** | | |
|---|---|---|---|---|---|---|
| | Fe | Cu | Pt | Fe | Cu | Pt |
| Thickness (Å) | 240±5 (242±5) | 50±3 (51±3) | 247±5 (246±5) | 268±4 (266±5) | 53±4 (52±3) | 222±6 (225±4) |
| Roughness (Å) | 7±2 (6±2) | 9±2 (8±2) | 5±1 (6±2) | 4±1 (5±1) | 11±2 (9±2) | 4±1 (5±2) |
| SLD ($10^{-6}$ Å$^{-2}$) | 59.1±2.2 (7.89±0.25) | 62.1±2.4 (6.30±0.16) | 139.1±3.9 (6.35±0.19) | 60.2±2.0 (7.88±0.27) | 61.1±2.3 (6.28±0.20) | 140.0±4.2 (6.36±0.23) |

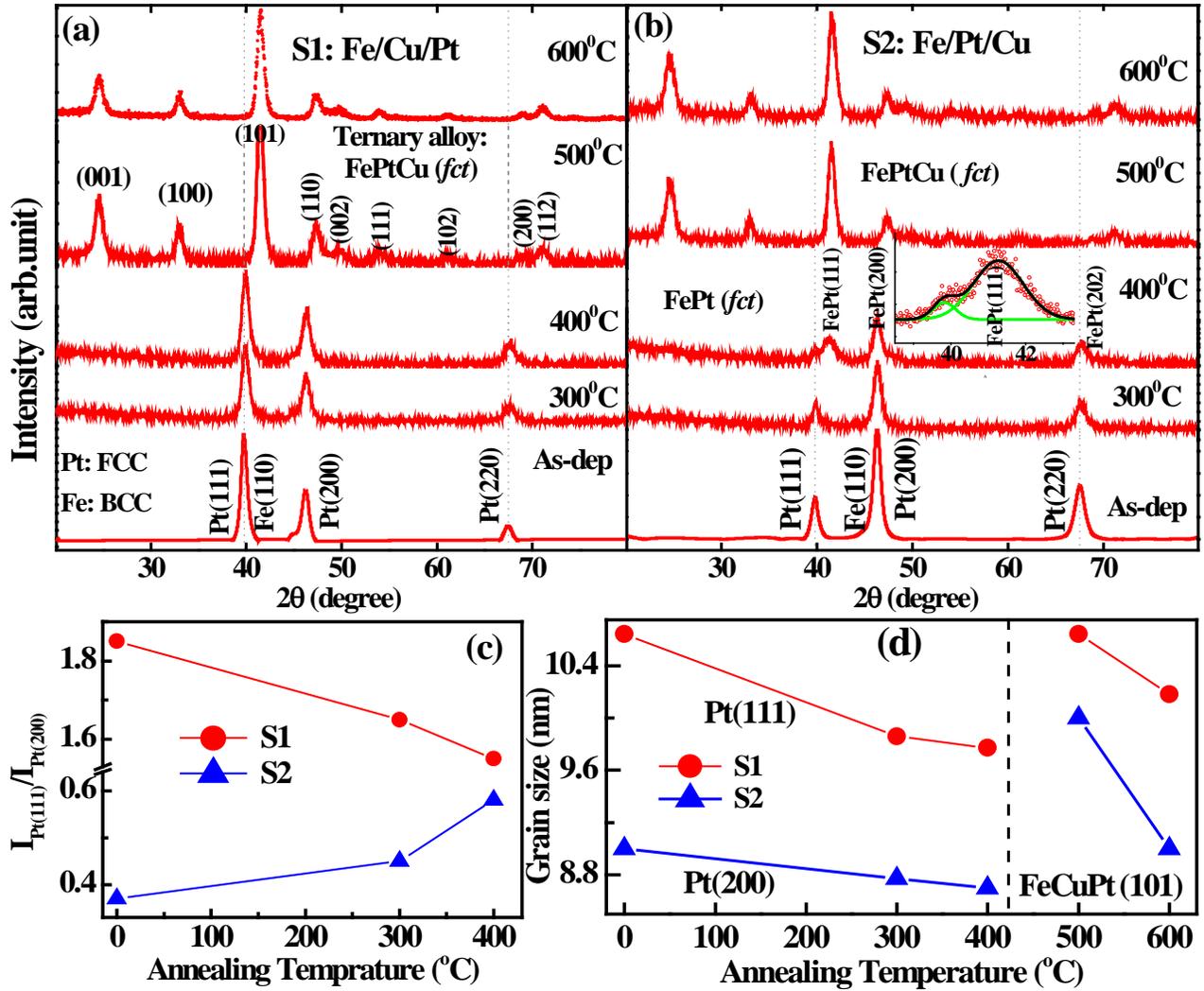

Fig. 1 GIXRD pattern from as-deposited and annealed heterostructures S1 (a) and S2 (b). Bragg reflections corresponding of different elements are indexed in the figure. On annealing S2 at 400 °C we obtained FePt alloy layer in addition to individual layers and the reflection from this alloy is also indexed. On annealing the heterostructures at 500 and 600 °C we obtained Bragg peaks corresponding to FePtCu ternary alloy, which are also indexed. (c) Intensity ratio for (111) and (200) Pt reflection as a function of annealing. (d) Variation of crystallite size of Pt and FePtCu as a function of annealing temperatures. We have considered the Bragg peak corresponding to Pt (111) for S1, Pt(200) for S2 and FePtCu (101) for S1 and S2, for estimating the crystallite sizes of these elements.

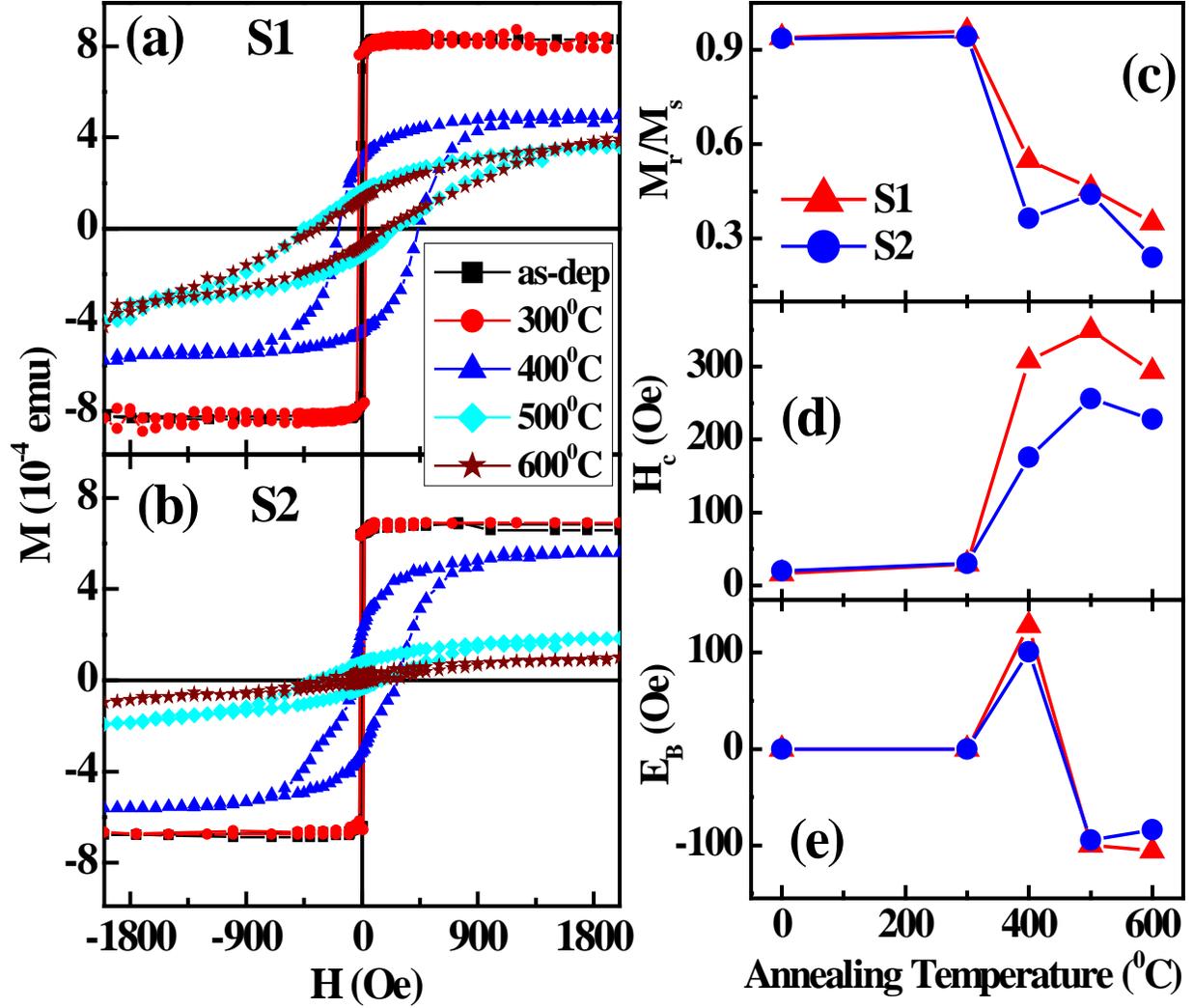

Fig. 2: Room temperature hysteresis curves ($M(H)$) from as-deposited and post annealing of heterostructures S1 (a) and S2 (b) at different annealing temperatures. (c), the ratio of remanent magnetization ($M_r$) to $M_s$, ($M_r/M_s$), (d) coercivity ($H_c$) and (e) variation of Exchange bias ($E_B$) as a function of annealing temperature for S1 and S2. Zero annealing temperature corresponds to As-deposited (as-dep) condition.

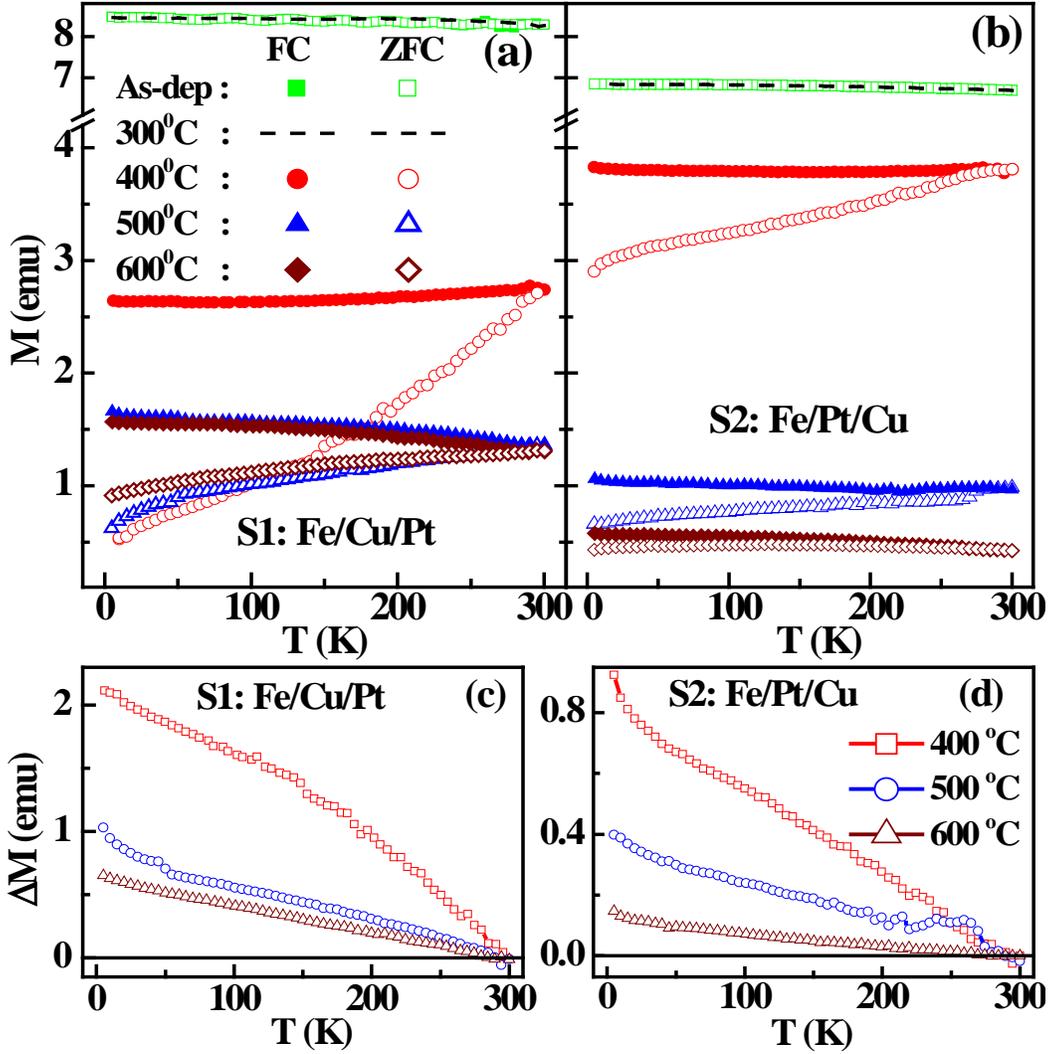

Fig. 3 : Magnetization (M) as a function temperature from as-deposited (as-dep) and post annealing of heterostructures S1 (a) and S2 (b) at different annealing temperatures for FC and ZFC condition. Difference between FC and ZFC magnetization data as a function of temperature from annealed heterostructures S1 (c) and S2 (d) at different temperatures 400 -600 °C.

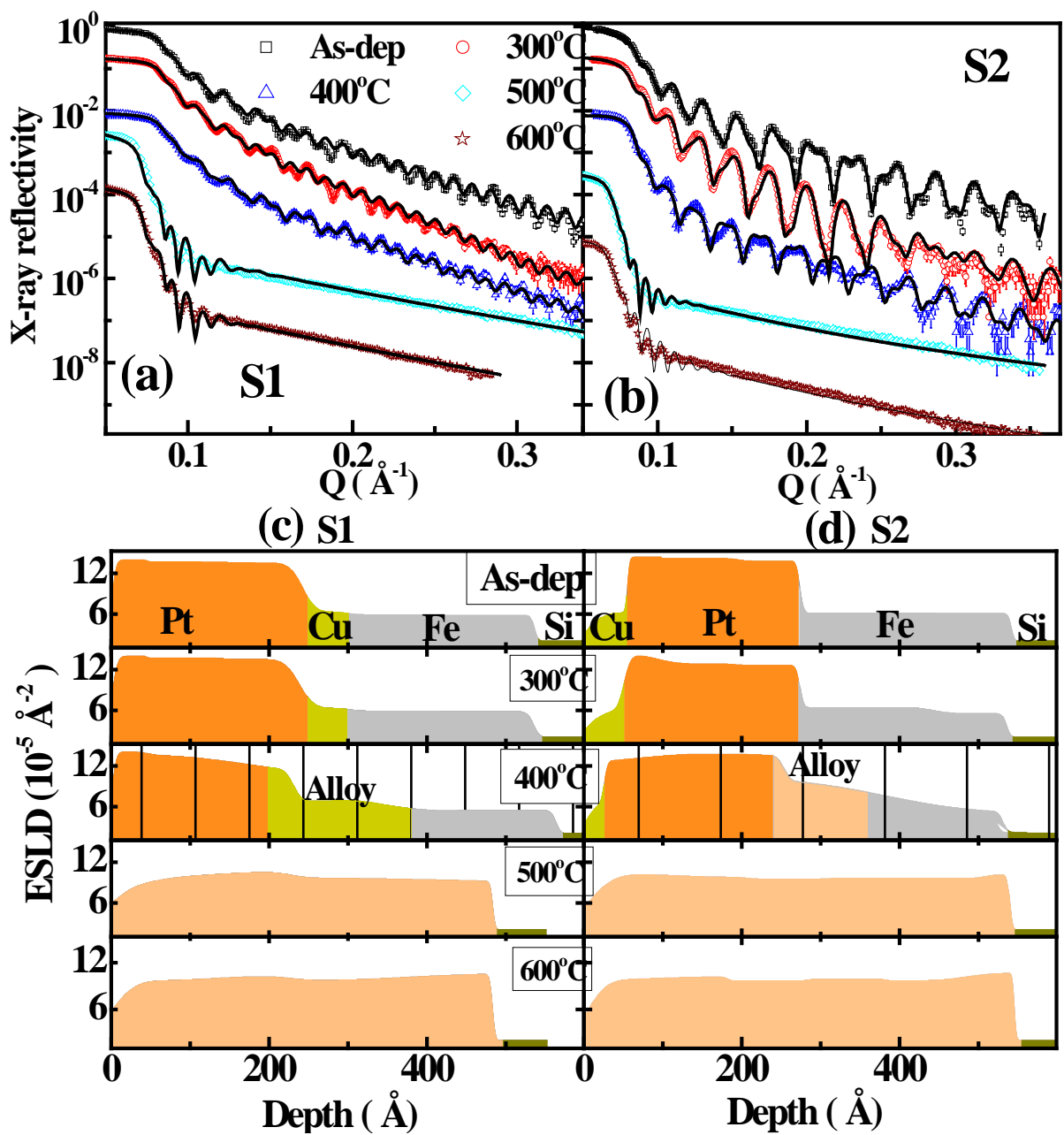

Fig. 4: X-ray reflectivity (XRR) data from heterostructures S1 (a) and S2 (b) for as-deposited and post annealing at different temperatures. (c) and (d) The electron scattering length density (ESLD) depth profile of S1 and S2, respectively, which best fitted (solid lines) the XRR data in (a) and (b).

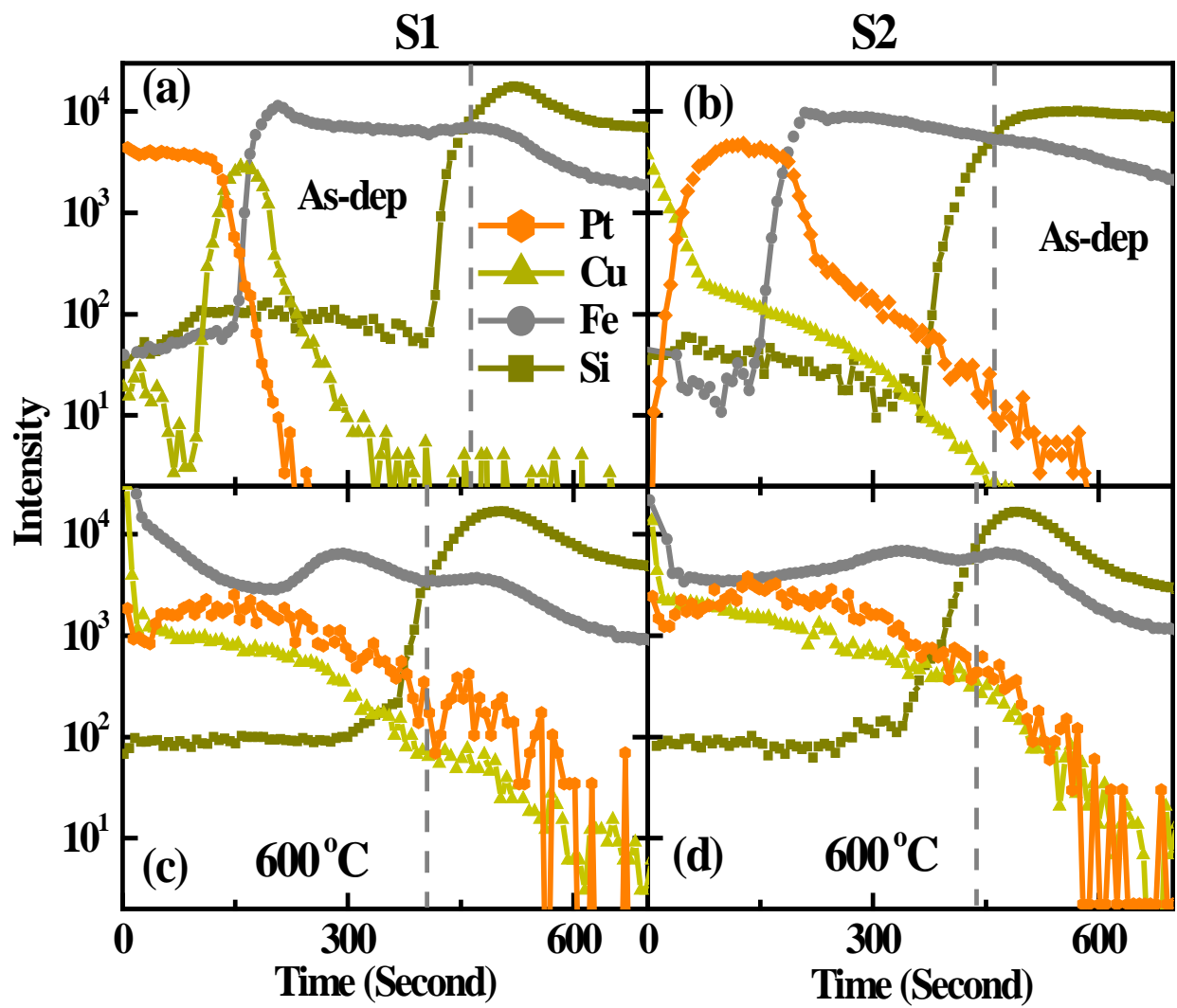

Fig. 5: Secondary ion mass spectrometry (SIMS) data from as-deposited heterostructures S1 (a) and S2 (b). SIMS data from annealed heterostructures S1 (c) and S2 (d) at 600 $^{o}$C.

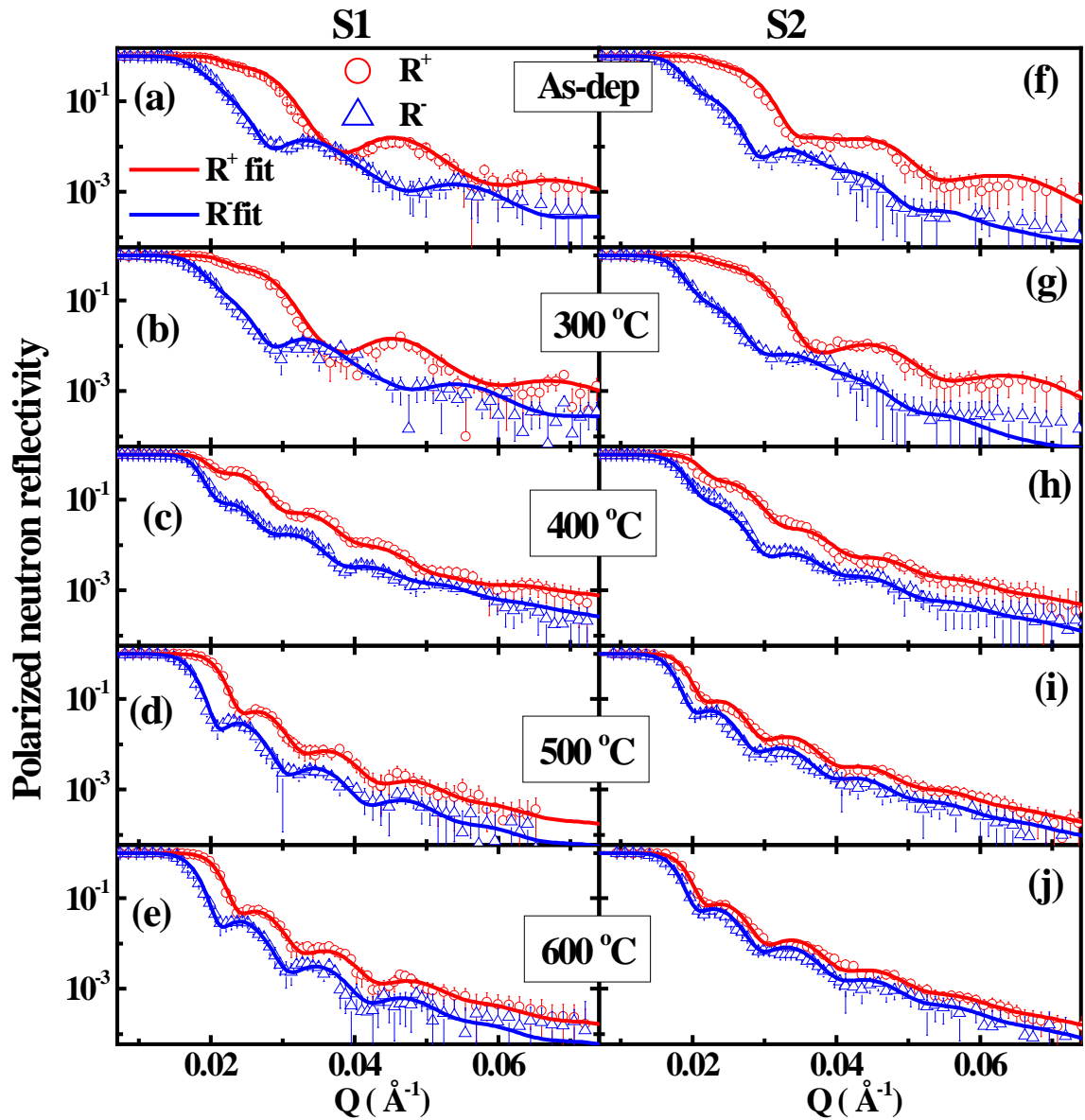

Fig.6: (a) - (e) Polarized neutron reflectivity (PNR) data from as-deposited (as-dep) S1 and S1 annealed at different temperatures. (f)-(j) PNR data from as-deposited S2 and S2 annealed at different temperatures.

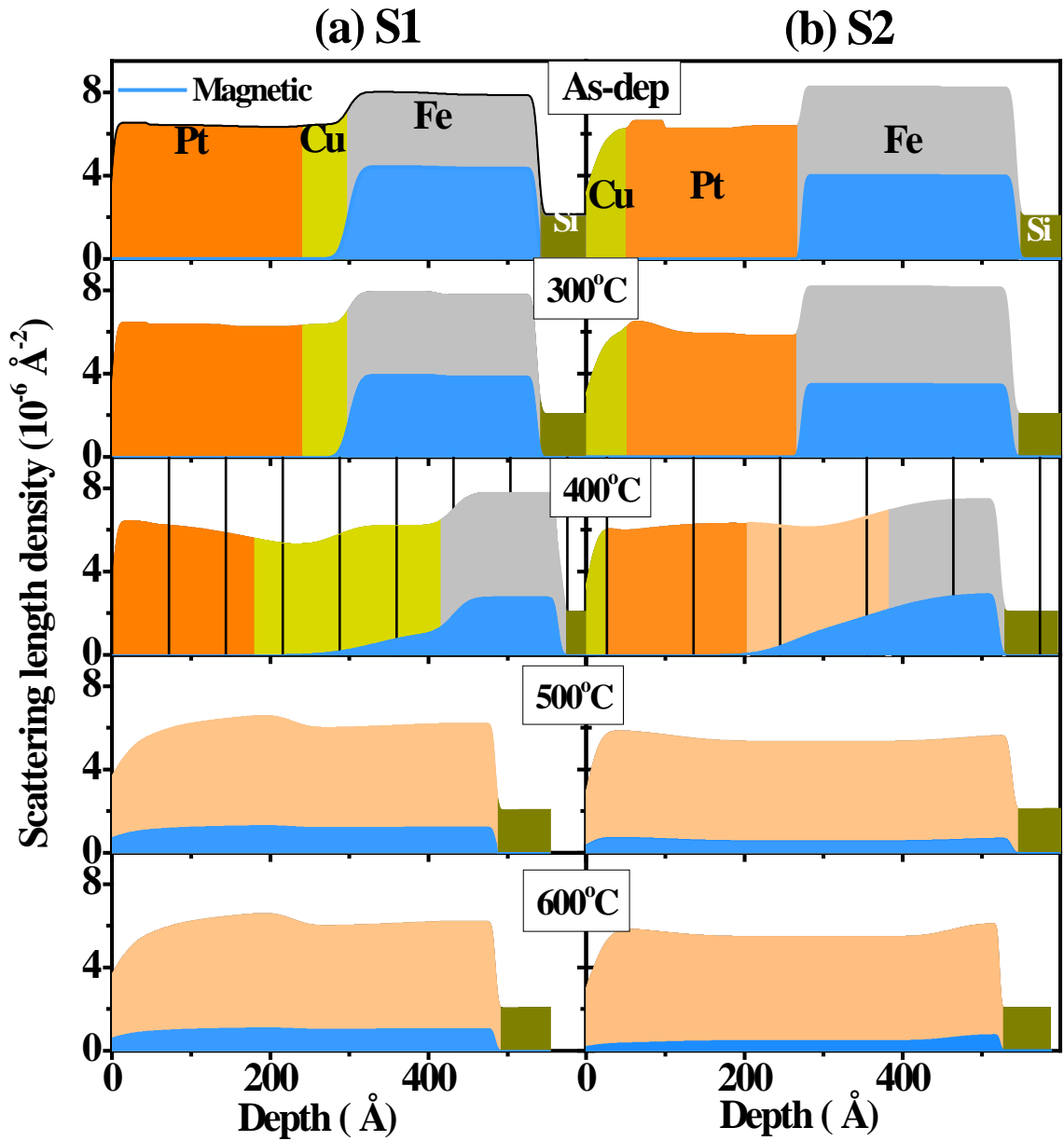

Fig. 7: Nuclear scattering length density (NSLD) and magnetic scattering length density (MSLD) (blue shaded region) depth profiles of heterostructures S1 (a) and S2 (b) for as-deposited and annealing at different temperatures. These profiles best fitted (solid lines) the corresponding PNR data in Fig. 6 (a) and (b).

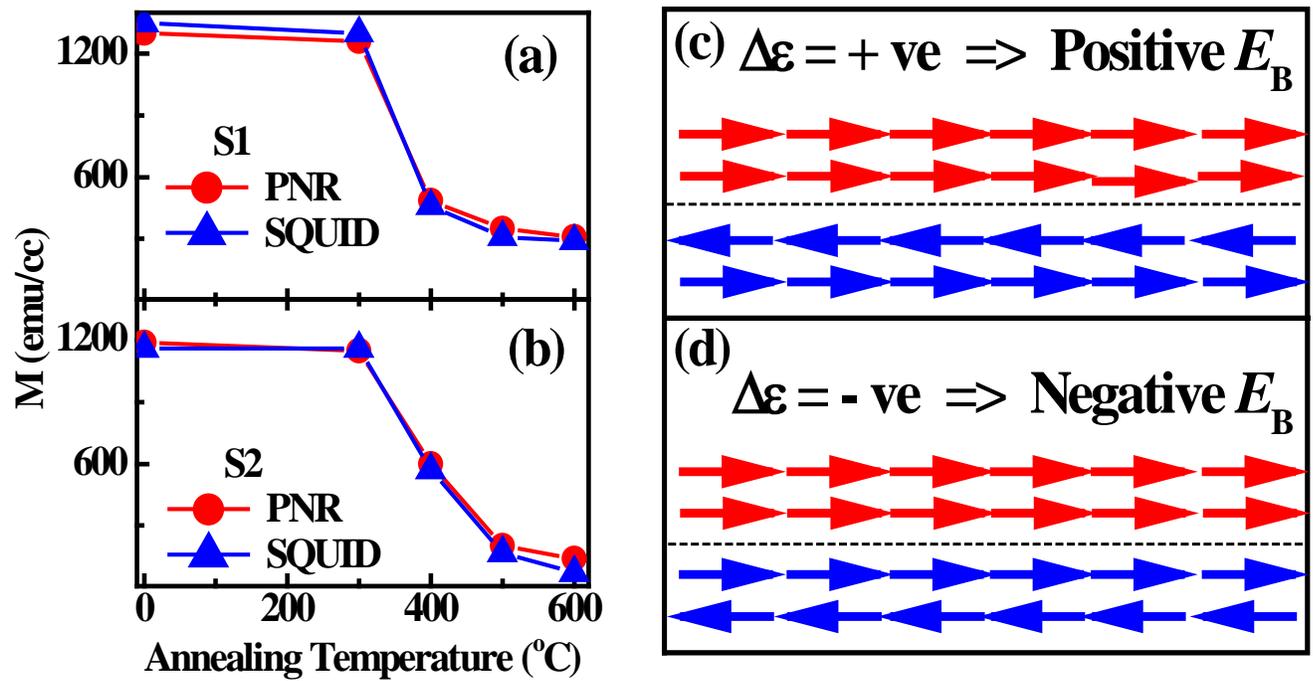

Fig. 8: (a,b) Comparison of magnetization of the heterostructures measured by SQUID and PNR (thickness weighted magnetization) measurements as a function of annealing temperature. (c, d) representation of spin configuration at the interface assuming antiferromagnetic/ferromagnetic coupling.

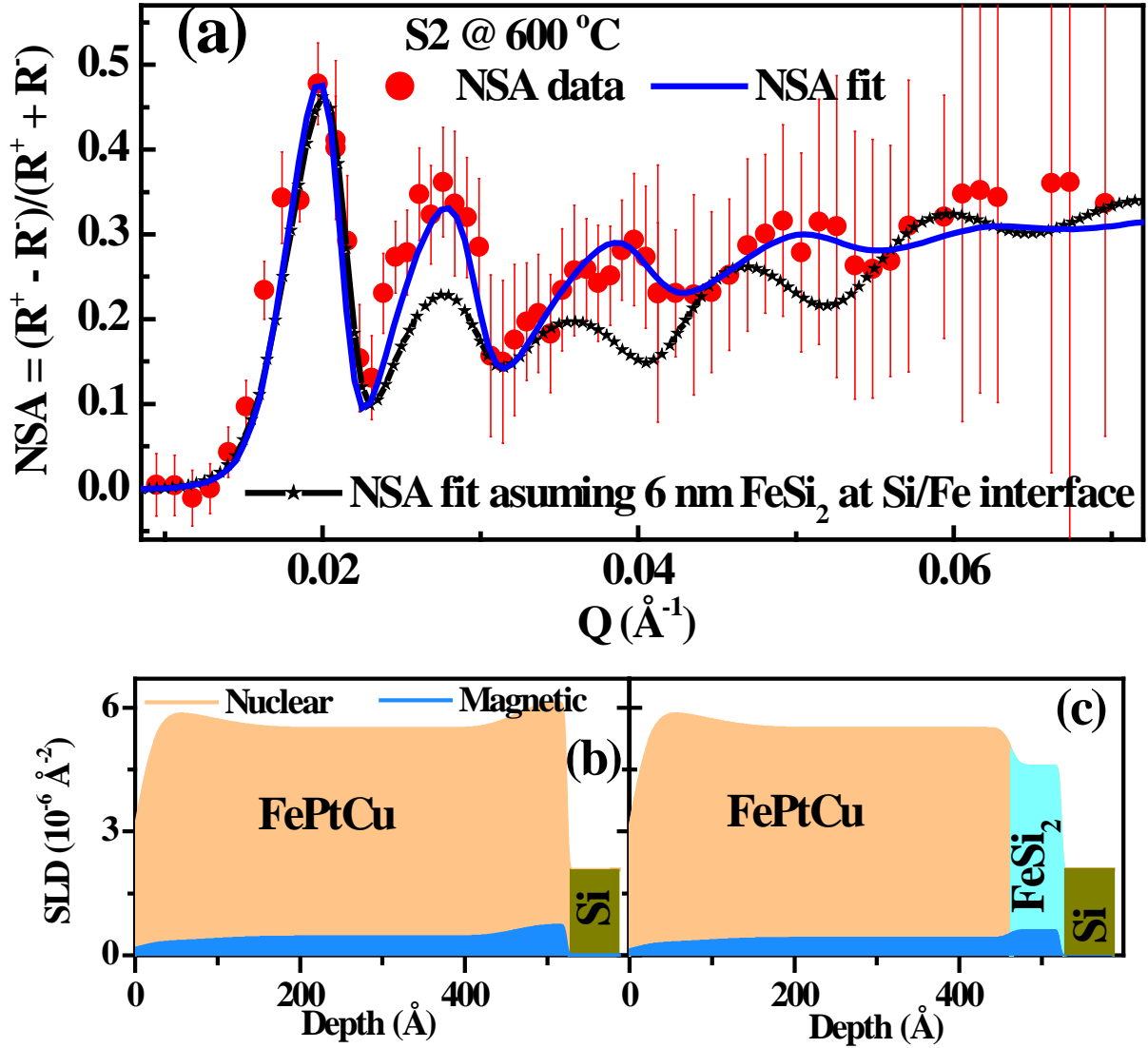

Fig. 9: (a) Normalized spin asymmetry (NSA = $(R^+ - R^-)/(R^+ + R^-)$) data (•) from heterostructure S2 on annealing at 600 °C. (b) Nuclear and magnetic scattering length density (NSLD and MSLD) depth profile of the alloy phase formed on annealing S2 at 600 °C, which best fitted (solid line, blue) the PNR (NSA) data in (a). (c) NSLD and MSLD depth profiles assuming a $FeSi_2$ layer at the substrate film interface which gave a NSA profile (line with star, black) shown in (a), suggesting this profile does not fit PNR data in the almost whole Q range.

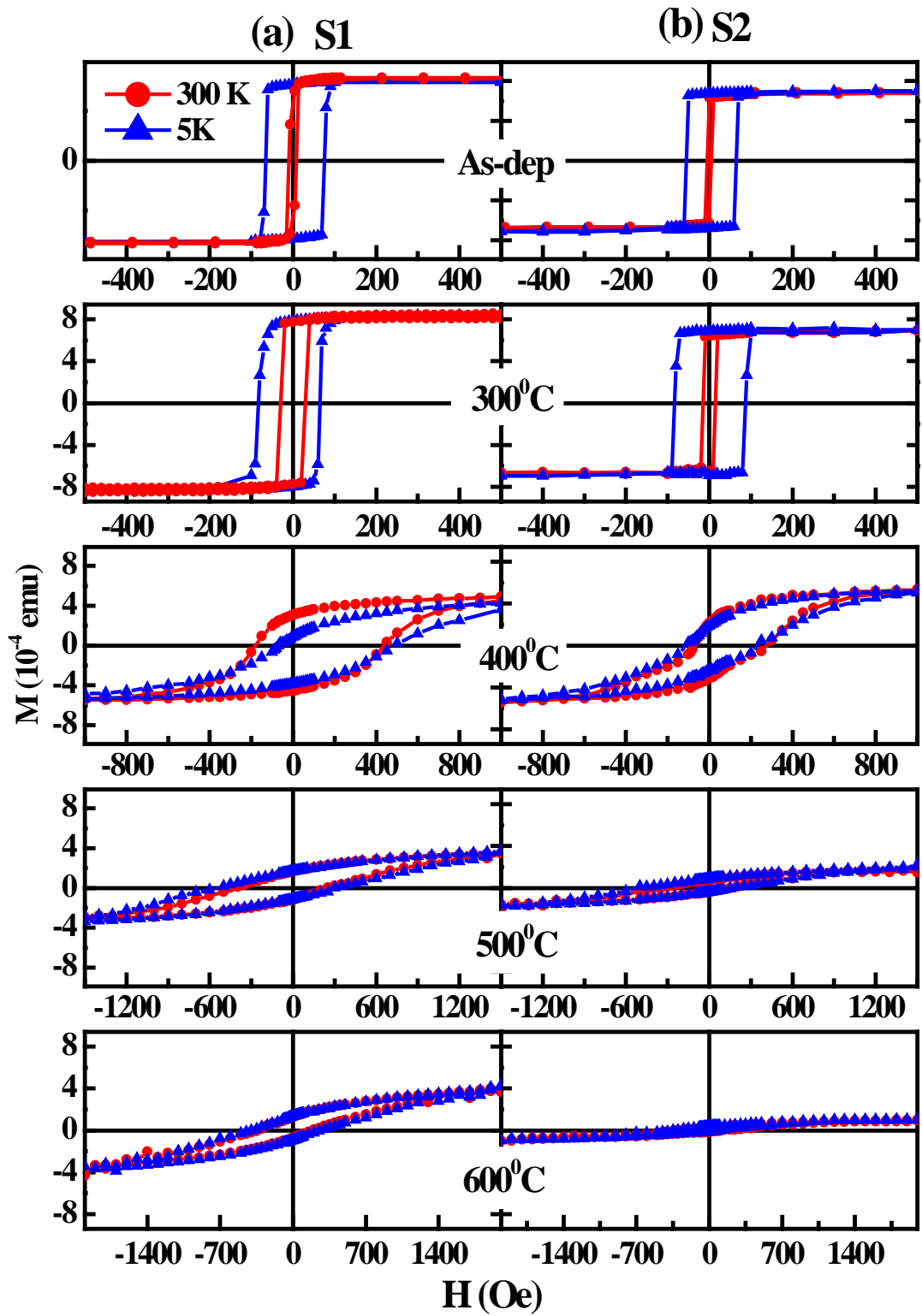

Fig. 10: Comparison of magnetization M (H) curves at 300 K and 5 K from heterostructures S1 (a) and S2 (b) for as-deposited and annealing at different temperatures conditions.